\documentstyle[aps,twocolumn]{revtex}
 \font\sf=cmss12  \font\gital=cmti10 scaled \magstep3    
    \def\U{\hbox{{\gital u$\,$}}} 
    \def\Vau{\hbox{{\gital v$\,$}}} 
 \def\MB{\hbox{\sf B}}       
 \def\MD{\hbox{\sf D}}       \def\MA{\hbox{\sf A}}
 \font\log=logo10 scaled \magstep2 \def\MA{\hbox{\log A}}
 \def\({\left(}       \def\){\right)}  
 \def\lk{\,\left[ \,} \def\rk{\,\right] \,} 
 \def\lw{\left\langle} \def\rw{\right\rangle}
 \def\wu#1{\sqrt{{#1} \,}^{ \hbox to0.2pt{\hss$ 
     \vrule height 2.5pt width 0.6pt depth -.5pt $} \;\! }}
 \def\vc#1{{\buildrel _\rightharpoonup \over #1}}
 \def\vcsm#1{ \def\sm{\raise 1.6pt\hbox to 5pt{\hss $_#1$}} 
          {\buildrel {_\rightharpoonup} \over \sm} \>\!\! }
 \def\nz{\par \noindent}      \def\cl#1{{\cal #1}}
 \def\schl#1{\widetilde{#1}}  \def\ov#1{\overline{#1}}   
 \def\0{\over }  \def\6{\partial }  \def\bu{_\bullet}  
 \def\ueb#1#2{{\;\buildrel{#1}\over{#2}\,}}
 \def\ln{\hbox{ln}}  \def\P{ {\mit\Pi} }
 \let\a=\alpha   \let\b=\beta   \let\d=\delta  
 \let\D=\Delta   \let\G=\Gamma  
   \def\eq#1{(\ref{#1})}      \def\nonu{\nonumber}
   \def\be#1{\begin{equation} \label{#1}}
   \def\ee{\end{equation}}
   \def\bea#1{\begin{eqnarray} \label{#1}}
   \def\eea{\end{eqnarray}}
\def\bsum{\hspace{.3cm}\raise 6.5pt\hbox{$\circ
          $}\hspace{-.45cm}\sum}
\def\btsum{\hspace{.05cm}\raise 5pt\hbox{$\circ
          $}\hspace{-.37cm}\sum}
\def\bosum{\hspace{.06cm}\raise 4pt\hbox{$\circ
          $}\hspace{-.33cm}\sum}
\def\st{{\hbox{\boldmath$*$}}}
\def\sst{\hbox{\scriptsize \boldmath$*$}}

\begin{document}

\title{Skeletons and Variation}   

   \renewcommand{\thefootnote}{\fnsymbol{footnote}} 
\author{Hermann Schulz$\,$\footnote[2]{} \\[-2.4cm] 
   \rightline{\vtop{\hbox{hep-ph/98$\,$08$\,$339} \vskip .1cm
   \hbox{TFT$\,$98 Regensburg}}} \vspace*{1.7cm} } 
\address{Institut f\"ur Theoretische Physik, Universit\"at 
   Hannover \\ Appelstra\ss e 2, D-30167 Hannover, Germany} 
\maketitle
   \renewcommand{\thefootnote}{\fnsymbol{footnote}} 
\footnotetext[2]{~Electronic adress~: 
                 ~hschulz@itp.uni--hannover.de} 

\begin{abstract}
Well known from the sixties, the pressure of e.$\!\;$g. 
massless $\phi^4$ theory may be written as a series of 
2PI-diagrams (skeletons) with the lines fully dressed. 
Varying the self--energy $\P$ in this expression, it turns 
into a functional $\U [Y]$ having a maximum in function 
space at $Y=\P$. There is also the Feynman--Jensen thermal
variational principle $\Vau [S]$, a potentially
non--perturbative tool. Here actions $S$ are varied.
   \par 
It is shown, through a few formal but exact steps, that
the functional $\U$ is covered by \Vau. The corresponding
special subset of trial actions is made explicit.
\end{abstract}
\narrowtext  \vspace*{.3cm}

\section{Introduction}

Beyond perturbation theory, we are in search for some
optimization calculus. The thermal variational principle,
while well appreciated in non--relativistic quantum
statistics, plays a merely minor role in field theory so
far. At present, there are two apparently different such
principles, which we shall call the Luttinger--Ward
principle \cite{luwa} and the Feynman--Jensen principle
\cite{feyn,vari}. For the study of the possible difference
of these two principles we shall concentrate on scalar
$\phi^4$ theory in the first two sections.

In 1960 Luttinger and Ward \cite{luwa} made a remark in
parentheses, that the pressure $p$ of their fermionic
system, after the diagram lines were fully dressed,
becomes minimal under variation of the self--energy. For
the massless thermal $\phi^4$--system with Lagrangian
$\,\cl L = (\6 \phi )^2/2 - g^2 \phi^4 /24\;$ this property
may be recapitulated as follows. The skeleton version for
the pressure $p$, which is $-1/V$ times the free energy
$F$, reads
\bea{p}
   p & = & - {1\02} \sum_P \,\ln \( { \P(P) - P^2 } \)
 + {1\02} \sum_P \; { \P(P) \0 \P(P) - P^2 } \nonu \\
 & & {} + {1\0\b V} \, \G  \qquad \hbox{with} \;\;\; 
  \G \,\equiv\, \sum_{n=1}^\infty f_n^{\rm 2PI} \quad .
\eea 
$\P(P)$ is the exact self--energy, $\sum_P = T\sum_n 
(2\pi)^{-3} \int\! d^3p\,$ is the thermal sum--integral, 
and $f_n^{\rm 2PI}$ are the 2--particle irreducible
contributions of order $g^{2n}$ to the logarithm 
$\ln (Z_{\rm int})$ of the partition function$\,$:
\bea{fn}
    \G  & = & 
    3 \hbox{\unitlength1cm  \begin{picture}(.85,.5)
      \put(.33,.12){\circle{.4}\circle{.4}} 
   \end{picture} }
 + 12 \hbox{\unitlength1cm  \begin{picture}(.65,.5)
      \put(.33,.12){\circle{.4}} \put(.5,.12){\circle{.4}}
      \end{picture} }
 + 2\cdot 12^2 \hbox{\unitlength1cm \begin{picture}(.85,.3)
      \put(.3,.02){\circle{.4}\circle{.4}}
      \put(.51,.34){\circle{.4}}
      \end{picture} }  \nonu \\[10pt]
  & & {} + \; {3\02} \cdot 12^3
      \hbox{\unitlength1cm \begin{picture}(1,.5)
      \put(.34,-.02){\circle{.4}\circle{.4}} 
      \put(.34,.38){\circle{.4}\circle{.4}}
      \end{picture} }
 + 6 \cdot 12^3 \hbox{\unitlength1cm
   \begin{picture}(1.1,.5)     \put(.45,.15){\circle{.6}}
   \put(.88,.15){\circle{.32}} \put(.76,.15){\circle{.6}}
   \end{picture} }  + \ldots \quad .
\eea 
The result \eq{p} is found in \cite{bloch,lawe}. It can 
be derived by a Legendre transformation \cite{cjt,freed}
or, equivalently, by minimizing the free energy 
\cite{rs}. It was recently taken up in \cite{pesh}, see 
also the article by A. Peshier in these proceedings.

Three minor modifications make the above expression 
\eq{p} to become the functional $\U$ of what we call
the Luttinger--Ward variational principle. We multiply 
\eq{p} with $\b V\,$. We supply the volume $V$ with 
periodic boundary conditions, hence the thermal sum turns 
into the ''bare sum'' $\,\btsum_P \equiv \sum_n 
\sum_{\vcsm p} = \b V \sum_P\,$. Third, we replace the 
self--energy $\P (P)$ by some function $Y(P)$ to be varied 
and introduce the notation $\,G(P) \equiv 1/ (Y(P)-P^2)\,$. 
Then, as is seen shortly, the principle states that
\bea{u}
  \U [Y] & = & {1\02} \bsum_P \,\ln \( G(P) \)
  + {1\02} \bsum_P \, Y(P) \, G(P) \nonu \\
  & & {} + \, \G [G]  \;\;\, \le \;\;\, 
    \b V\, p \quad .
\eea 
The fact that here the pressure is at maximum, in contrast 
to that of \cite{luwa}, is due to the boson content of our 
system. The replacement $\P \to Y$ had to be performed even 
in the last term, i.e. in the diagram lines of \eq{fn}.
So, these lines, the bones of the skeleton, have become 
variable propagators $G(P)$. 

For taking the functional derivative of $\U$ with respect 
to $Y(Q)$ at an arbitrary ''position'' $Q$, we appreciate
the general functional relation \cite{kapu} 
\be{krel}
   2\, G_0^2\; \d_{G_0} \, f = G\, \qquad\quad 
         \Big( \; f \equiv \ln \( Z \) \; \Big) \quad ,
\ee 
between bare and dressed lines, which is valid separately 
in each loop order \cite{rs}. Hence
\bea{df}
  2 G^2 \,\d_G \, \G [G] & = & G^{\rm 2PI}[Y] 
    \, = \hbox{\unitlength1cm  \begin{picture}(1.3,.3)
    \put(.1,.12){\line(1,0){.4}}
    \put(.65,.12){\circle*{.3}}
    \put(.8,.12){\line(1,0){.4}} 
    \put(.44,.36){\tiny 2PI} \end{picture}} \nonu \\[5pt]
  & = & \, - \, G^2\, \P^{\rm 2PI} [Y] \quad 
\eea 
and
\bea{du}
  2\, \d_{Y(Q)} \, \U & = & - G^2 \( Y-\P^{\rm 2PI}
  [Y] \) \,\; \ueb{!}{=} \,\; 0 \nonu \\[5pt]
  & \Rightarrow & \qquad
  \quad Y(Q) = \P (Q)  \quad ,
\eea 
because the full $\P (Q)$ (up to a given order $n$) is
re\-con\-struc\-ted by iterative use of $Y_m
= \P_m^{\rm 2PI} \,$ with $m<n$ \cite{rs}. At any position
$Q$, when $Y$ increases there, $\d_{Y(Q)} \U$ reaches the
zero coming from positive values and turning into negative.
So, the functional $\U$ has a true maximum at $Y=\P\,$.

Some warning is in order, if the sum $\sum f_n^{\rm 2PI}$ 
in $\U$ is truncated. Let $2 n^\prime$ be the highest 
$g$--power contained in $\U\,$. Then, at best, the
resulting $\P$ would be correct to order $g^{n^\prime + 3}$
only ($n^\prime \ge 4$). $\P$ in $g^{12}$, for instance,
needs $\U$ to be developed to $g^{2n^\prime} = g^{18}$.
This unfortunate fact is due to reduction of $g$--order by
evaluation. Things are realized by power counting of soft
scale contributions, or along the lines given by Braaten
and Nieto \cite{bn}. One might question for a better
functional, which is free from this defect. For the
following we stay with the exact expression \eq{u}.

To introduce the Feynman--Jensen variational principle we
must distinguish between the action $S$ of a trial theory
and the action $S\bu$ of the true $\phi^4$--system under
study. With or without a bullet index, we have by
definition $S = -\int^\b \cl L\,$, $\int^\b = \int_0^\b
\! d\tau \int\! d^3r\,$ and $\b = 1/T\,$. Then \cite{feyn}
\bea{z}
  Z\bu & = & \int\! \cl D \phi \; e^{-S\bu}
       = \( \int\! \cl D \phi \; e^{-S} \) \cdot
         { \int\! \cl D \phi \; e^{-S} e^{S -S\bu} \0
           \int\! \cl D \phi \; e^{-S} } \nonu \\[5pt]
    & = & Z \lw e^{S-S\bu} \rw 
    \,\;\ge \,\; Z e^{\lw S-S\bu \rw} \quad .
\eea 
with $Z$, $Z\bu$ the partition functions of trial and 
studied theory, respectively. The inequality in \eq{z} is 
the familiar Jensen inequality \cite{blat} applied to
functional integrals. The functional measure is obviously
irrelevant here. Note that the average refers to the trial 
action $S$, which usually might be the simpler theory. 
Taking the logarithm of \eq{z} we arrive at
\be{fj}
  \Vau [S] \; = \;\ln \( Z \) + \lw S-S\bu \rw \; 
             \le \; \ln \( Z\bu \) \quad .
\ee 
Also the functional $\Vau$ has a true maximum, this time
at $S=S\bu\,$. The principle \eq{fj} was studied with 
detail in \cite{vari} and applied to gauge fields. 
Working with free trial theories, several known results 
could be reproduced in \cite{vari}. But to get more 
information from the principle, the free trial space 
has turned out being too poor.

   
\section{The way $\U$ is contained in \Vau }

In three steps the principle $\U$ will be reformulated 
until it has the structure of $\Vau\,$. 

Step one. To get rid of the infinite sum $\G$ in \eq{u}, 
we exploit the fact that the l.h.s. of \eq{u} regains a 
physical meaning by the replacements $Y G \to \P G$ and 
$\U \to \ln\(Z\)\,$. Here, $Z$ is the partition function 
of a $\phi^4$ system with some momentum--dependent mass 
$G_0^{-1} + P^2$ {\sl such that} a given exact propagator 
$G=1/(Y-P^2)$ turns out perturbatively. Note that 
$\P = G^{-1} -G_0^{-1}$ differs from $Y=G^{-1}+P^2\,$. 
Since $\G [G]$ is a functional of $G$ only, all the bare 
lines, which constitute the partition function $Z$
diagrammatically, must be re--expressed by $G\,$:
\bea{gamma}
  \sum_{n=1}^\infty f_n^{\rm 2PI} & \equiv & 
   \; \G [G]  \nonu \\[2pt]
   & = & \lk \ln \(Z\) 
  - {1\02} \bsum \ln \( G \) - {1\02} \bsum \P G 
  \right]_\st  \nonu \\[4pt]
  & & \hspace*{-1.2cm} = \; 
      \ln \( Z_\st \) - {1\02} \bsum \ln \( G \)
    - {1\02} \bsum \( 1 - {G \0 G_{0\,\st}} \) \;\; .
\eea 
Without star index, the inner line of \eq{gamma} is nothing
but the skeleton formula \eq{p}, but this time for the
massive system just introduced. The prescription $\st$ has 
the following definite meaning$\,$: solve the relation 
\eq{krel}, i.e. $ 2 G_0^2 \d_{G_0} f = G\,$, for $G_0\,$; 
then replace all $G_0$'s in the square bracket by the 
resulting functional $G_0 [G] \equiv G_{0\,\st}\,$. This 
is a Legendre transformation \cite{cjt,freed}. As $G$ does 
not change under the \st--operation, the second term reached 
the second line unchanged. But it will now be canceled when 
using \eq{gamma} in \eq{u}$\,$:
\bea{10}
  \U [Y] & = & {1\02} \bsum Y G + \ln \( Z_\st \) 
    - {1\02} \bsum \( 1 - {G \0 G_{0\,\st}} \)
       \nonu \\[5pt]
    & \le & \,\; \b V\, p \; = \; \ln \( Z\bu \) \quad .
\eea 
The inequality refers to variation of $Y$. While still 
$G=1/(Y-P^2)\,$, the objects $G_{0\,\st}$ and $Z_\st$ are 
some nontrivial functionals of $G$, hence of $Y\,$.

Step two. We transcribe the inequality \eq{10} into
functional integral language. Exponentiating we have
\bea{11}
   e^{{1\02} \,\bosum \( YG-1+{G\0 G_{0\sst} }\) }\; Z_\st 
   & \;\le\; & Z\bu \quad , \quad \nonu \\[5pt]
   e^{{1\02} \,\bosum \( Y G - 1 + {G\0 G_{0\sst}} \)}
   \int \cl D \phi \; e^{-S_\st}    
   & \;\le\; & \int \cl D \phi \; e^{-S\bu } \quad ,
\eea 
where 
$S\bu = - \int^\b \!\( \cl L_0 + \cl L_{\rm int} \)\,$, 
but $\, S_\st = - \int^\b \! \big( \cl L_{0\,\st} + $ $
      + \cl L_{\rm int} \big)\,$ 
with
\bea{12}
     \cl L_{0\,\st} & = &\cl L_0 |_{G_0 \to G_{0\sst}}
    \quad , \nonu \\[5pt]
    - \int^\b \cl L_{0\,\st} 
    & = & {1\02} \sum_P \schl \phi (-P) 
      { 1 \0 G_{0\st} } \schl \phi (P) 
     \; \equiv \; S_{0\,\st} \quad . 
\eea 

We turn to the exponential in \eq{11}, which might be 
some functional average over two fields $\schl \phi\,$. 
In fact, if we first separate a factor $G$ by
\be{13}
  YG - 1 + {G \0 G_{0\sst}} \; = \; 
  G \( {1\0 G_{0\sst} } - {1\0 G_{00}}\)\quad , 
\ee 
where $G_{00} \equiv -1/P^2$, and write this factor as
\bea{14}
  & & \hspace{-.5cm}
  G(P) 2 G_{0\,\st}^2 \,\d_{G_{0\sst}} \ln \( Z_\st \)
      \; =  \nonu \\[8pt]
   & = & { 2 G_{0\,\st}^2 \0 Z_{\st}} \,\d_{G_{0\sst}}
     \int\! \cl D \schl \phi \; e^{- {1\02\b V} \bosum_Q
        \schl \phi (-Q) {1\0 G_{0\sst}} \schl \phi (Q) } 
        e^{-S_{\rm int} } \quad \nonu \\[5pt]
   & = & {1\0 \b V} \lw \schl \phi (-P) \,\schl 
        \phi (P) \rw_\st \quad ,
\eea 
then the exponential becomes
\bea{15}
 & & \hspace*{-.8cm}
    {1\02} \bsum \( Y G - 1 + {G\0 G_{0\sst}} \) \; = 
       \nonu \\[5pt]
    & = & {1\0 2} \sum_P \lw \schl \phi (-P) 
     \( {1 \0 G_{0\sst} } - {1 \0 G_{00} } \)
     \schl \phi (P) \rw_\st  \nonu \\[5pt]
    & = & \lw\, S_{0\,\st} - S_{0\,\bullet}\,\rw_\st 
    \; = \; \lw\, S_\st - S\bu \,\rw_\st \quad .
\eea 
Using \eq{15} in \eq {11}, the result
\be{16}
  Z_\st \; e^{ \lw S_\st - S\bu \rw_\st } \;\; 
  \le \;\; Z\bu \; = \;
  Z_\st \; \lw e^{S_\st - S\bu} \rw_\st \quad 
\ee 
formally agrees with \eq{z} and is a Jensen inequality. 

Step three is merely one in mind rather than in 
formulation. So far we varied functions $Y$ around $\P$ 
or, equivalently, $G$'s around $1/(\P-P^2)\,$. But this is 
not the variation in the Feynman--Jensen functional. The 
latter depends on actions. So, we now change the 
philosophy and consider $G_{0\,\st}\,$ to be the variable
function. Hence \eq{16} becomes the Feynman--Jensen 
principle, indeed. But the class of trial actions $S_\st$ 
is very restricted. The full interaction $\cl L_{\rm int}$ 
is part of $S_\st$ and remains untouched. Due to \eq{12} 
$S_\st$ differs from $S\bu\,$ only in some variable 
momentum--dependent mass term. Moreover, the solution to 
the optimization problem has become trivial$\,$: the 
equality sign in \eq{16} is simply reached at vanishing 
mass term. The true value of using $\Vau$, in contrast, 
is in optimizing approximations for a physics too hard 
for an exact or even perturbative analysis.

From the fact, that we are on a very safe ground with 
the special Feynman--Jensen variational principle \eq{16}, 
we are led to reverse the order of the above three steps. 
Then they guide a possible derivation of the skeleton 
formula \eq{p} including its variational property \eq{u}. 
This idea is followed up in the next section. 


\section{The skeletons of Yang and Mills} 

We derive the skeleton formula for the pure gluon system 
by following the above equations in backward direction. In 
covariant gauges the action includes a gauge fixing term, a 
term arising from the Faddeev--Popov determinant and the
3-- and 4--point interactions, 
\be{17}
 S\bu = S_0 + S_{\rm g.f.} + S_{\rm FP} [A] 
        + S_{\rm int}  \quad .
\ee 
For the functional inequality \eq{16} to make sense (but 
of no relevance in the sequel), $S_{\rm FP}$ must be 
viewed as a functional of the fields $A_\mu^a$ 
\cite{vari}. The trial action $S_\st$ and the action
$S\bu$ have the same gauge fixing parameter $\a$ (see
\S~II.B of \cite{vari}). Then, the only difference 
between the two actions is a momentum--dependent mass 
term, which we immediately write as 
\bea{18}
  & & \hspace{-.4cm} S_\st - S\bu \; =
      \nonu \\[4pt]
  & = & \; {1\02} \sum_P \schl A_\mu^a (-P) 
  \lk {1\0 G_{0\,\st}} - {1\0 G_0 } \right]^{\mu\nu} 
  \schl A_\nu^a (P) \quad .
\eea 
The fully dressed Greens function $G_{\mu\nu}$ of 
the $\st$--theory enters when averaging \eq{18} 
\bea{19}
  & & \hspace*{-.4cm} 
  \lw \, S_\st - S\bu \,\rw_\st \; =
      \nonu \\[4pt]
  & = & \; {n\02}\, \b V 
  \sum_P \lk {1\0 G_{0\,\st}(P)} - {1\0 G_0 (P) }
  \right]^{\mu\nu}  G_{\mu\nu} (P) \quad 
\eea 
with $n= C_A = N^2-1\,$. Introducing a variable matrix 
self--energy $Y^{\mu\nu}(P)$ we are led to
\bea{20}
  \lk {1\0 G_{0\,\st}} - {1\0 G_0 } \rk G   
  & = & \lk {1\0 G_{0\,\st}} - {1\0 G_0 } \rk  
  \( 1 \0 G_0^{-1} - Y \) \nonu \\[4pt]
  & = &  {1 \0 G_{0\,\st} }\, G - 1 - Y G \quad .
\eea 
with all products being Lorentz matrix multiplications.
Some signs differ compared to \eq{13} due to Minkowski
metrics $\; +---\;$ and the conventions of \cite{vari}. 
Changing from variable $G_{0\,\st}$ to variable $Y$, the 
logarithm of \eq{16} reads
\bea{21}
  \U[Y] & = & n \bsum_P X[G] \; - \; {n\02} \bsum_P Y G \; 
  + \; \G \lk G \rk  \nonu \\[5pt]
  & \le & \,\; \ln \( Z\bu\)  \quad 
\eea 
with
\bea{22} 
  \G [G] & = & \lk \ln \( Z\) 
  - n \bsum_P X[G] + {n\02} \bsum_P \( 
    {1 \0 G_0} G - 1 \) \right]_\st \quad  \nonu \\[4pt]
  & \ueb{?!}{=} & \;
  \;\; \sum_{n=1}^\infty f_n^{\rm 2PI} \quad . 
\eea 
So far, the term with $X[G]$ has been only added and 
subtracted. But now we require the square bracket to be 
the sum over 2PI contributions to $\ln \( Z_\st \)\,$. 
With the Yang--Mills counterparts of \eq{krel} and \eq{df}, as 
listed in \eq{23}, this condition turns into the
matrix differential equation in the third line$\,$:
\bea{23}
 2 G_0 \( \d_{G_0} f \) ) G_0 & = & n\, G 
      \quad , \nonu \\[5pt]
 {2\0n}\,\d_G \,\G [G] & = & {1\0 G_{0\,\st} } - {1\0 G} 
      \quad , \nonu \\[5pt]
 \d_G \,\bsum X [G] & = & {1\0G} \quad .
\eea 
The solution to the $X$ equation needs \eq{24} and 
is given (apart from some $Y$--independent constant) in 
\eq{YM} below$\,$: first term. Herewith, one might drop the 
question mark in \eq{22}. 

To supply the first two terms of \eq{21} with detail, we 
need the structure of the exact gluon pro\-pagator. Due 
to Weldon \cite{wel} this propagator and the self energy 
may be written as \nz
\bea{24}
  G^{\mu\nu} & = &\D_t \MA^{\mu\nu} + \D_\ell \MB^{\mu\nu} 
  + {\a \0P^2} \ov{\MD}^{\mu\nu} \quad , \nonu \\[5pt]
    Y^{\mu\nu} & = & Y_t \MA^{\mu\nu} 
  + Y_\ell \ov{\MB}^{\mu\nu} 
  + P^2 \( \MB^{\mu\nu} - \ov{\MB}^{\mu\nu} \) \quad ,
\eea 
where $\D_{t,\ell} = 1/( P^2 - Y_{t,\ell} )\,$. Apart 
from $A=g-B-D\,$ all the Lorentz matrices in \eq{24}
are dyadic products of the four--vectors $P=(P_0,\vc p)$
and $\schl P = (p,P_0 \vc p /p)\,$, or of $R=P-y \schl P$
and $\schl R = \schl P - y P\,$, respectively. To be 
specific$\,$: 
   $B      = - \schl P \circ \schl P / P^2 \, $, 
   $D      = P \circ P /P^2 \, $, 
   $\ov{B} = -\schl R \circ \schl R /P^2 \, $,
   $\ov{D} =  R \circ R /P^2 \, $. 
We emphasize that even the Weldon coefficient $y$ is 
varied. With \eq{24}, and after preparing the trace 
over all pairs of the above matrices, one verifies
$\, Y^{\mu\nu} G_{\nu \mu} = 2 Y_t \D_t + Y_\ell \D_\ell 
   - \a\, y^2 \, $ with ease. The Luttinger--Ward 
functional $\U$ for the gluon system then finally 
reads
\bea{YM}
  \U [Y] & = & {n\02} \bsum_P \ln \( 
        \left[ {1 \0 Y_t - P^2} \right]^2 \!
           {1 \0 Y_\ell - P^2 } \)  \nonu \\[4pt]
 & & \hspace*{-1cm} + \; 
     {n\02} \bsum_P \( {2 Y_t \0 Y_t - P^2} + 
     {Y_\ell \0 Y_\ell - P^2} + \a\, y^2 \) + \G \quad .
\eea 
The maximum is reached at $Y_t=\P_t$, $Y_\ell=\P_\ell$ and 
$y=b\,$, where the coefficient $b$ is defined by \eq{24} 
at $Y^{\mu\nu}=\P^{\mu\nu}$ and is given by $\wu 2 \P_d / 
\P_c$ in earlier notations \cite{flesh}. At maximum 
${1\0 \b V} \U$ becomes the pressure of the hot gluon 
medium. 


\section{Conclusions} 

The competition between the Luttinger--Ward and the
Feynman--Jensen variational principle is won by the latter. 
It contains the former as a special case, whose space of 
trial actions is reduced to a rather trivial variation of 
mass terms. Nevertheless, the machinery relating the two 
principles can be used to derive the skeleton formula for 
the pressure of both, the massless scalar $\phi^4$ theory 
and the gluon system. 

\nz
{\sl I thank Jens Reinbach for deriving the
 Luttinger--Ward functional \eq{YM} for Yang--Mills 
 fields along the lines given in \cite{rs}.} 


\end{document}